\documentclass{article}

\pdfoutput=1

\usepackage{authblk}
\usepackage{epsfig}
\usepackage{epstopdf}
\usepackage{natbib}
\usepackage{fp}
\usepackage[capbesideposition=right]{floatrow}
\usepackage{wrapfig}
\usepackage[hmargin=0.75in,vmargin=1in]{geometry}

\usepackage{amsmath}
\usepackage{amssymb}
\usepackage{hyperref}

\newcommand{\Rs}{$ R_{\odot} $}
\def\ion[#1 #2]{#1\,{\sc #2}}

\begin{document}

\title{Coronal Magnetic Field Topology From Total Solar Eclipse Observations}

\author[1]{Benjamin Boe}

\author[1]{Shadia Habbal}

\author[2]{Miloslav Druckm\"uller}

\affil[1]{\small Institute for Astronomy, University of Hawaii, Honolulu, HI 96822, USA}
\affil[2]{\small Faculty of Mechanical Engineering, Brno University of Technology, Technick 2, 616 69 Brno, Czech Republic}

\maketitle

\begin{abstract}
\normalsize Measuring the global magnetic field of the solar corona remains exceptionally challenging. The fine-scale density structures observed in white light images taken during Total Solar Eclipses (TSEs) are currently the best proxy for inferring the magnetic field direction in the corona from the solar limb out to several solar radii (\Rs). We present, for the first time, the topology of the coronal magnetic field continuously between 1 and 6 \Rs, as quantitatively inferred with the Rolling Hough Transform (RHT) for 14 unique eclipse coronae that span almost two complete solar cycles. We find that the direction of the coronal magnetic field does not become radial until at least 3 \Rs, with a high variance between 1.5 and 3 \Rs \ at different latitudes and phases of the solar cycle. We find that the most non-radial coronal field topologies occur above regions with weaker magnetic field strengths in the photosphere, while stronger photospheric fields are associated with highly radial field lines in the corona. In addition, we find an abundance of field lines which extend continuously from the solar surface out to several solar radii at all latitudes, regardless of the presence of coronal holes. These results have implications for testing and constraining coronal magnetic field models, and for linking {\textit{in situ}} solar wind measurements to their sources at the Sun.

\end{abstract} 

\section{Introduction} 
\label{intro}

The shape of the solar corona was first characterized during Total Solar Eclipses (TSE) with hand-drawn sketches and early photographic techniques (e.g. \citealt{Maunder1899} and ref. therein). It was found to vary in a periodic manner correlated with the sunspot cycle  \citep{Darwin1889} as observed on the solar surface \citep{Schwabe1844}. The later discovery that the sunspot cycle is in fact a magnetic cycle \citep{Hale1919}, implied that the variable shape of the corona was magnetic in nature. During solar minimum the corona was found to have large polar plumes \citep{Saito1958} that dominate the north and south poles, {which originate from polar coronal holes} \citep{Munro1972}. The equatorial regions are dominated by streamers \citep{Maunder1899} -- which tend to have a higher electron density, and thus a higher visible light coronal intensity \citep{vandeHulst1950}. During solar maximum the coronal structure was less organized, with mostly radial features (e.g. \citealt{Newkirk1967} and ref. therein). 


\par
In recent years, the most common technique employed to measure the direction of the coronal magnetic field is through linear polarization observations {of coronal emission lines} with a coronagraph {(e.g. {\citealt{Habbal2001B,Gibson2017, Dima2019}})}. {These} observations have remained somewhat limited in extent however, as ground-based coronagraphic observations {of polarized emission lines} can only probe below $\approx 1.5$ {solar radii (\Rs)} due to scattering {of sunlight by the Earth's atmosphere}. Polarization observations {of coronal emission lines} also have to contend with the `Van Vleck' angle ambiguity \citep{vanVleck1925}, where structures curved more than $54.7^{\circ}$ with respect to the radial direction from Sun center will have their polarization angles rotate by 90$^\circ$ from the true magnetic field direction. This ambiguity can lead to misleading identification of the field direction in highly curved structures such as in active regions and the base of streamers.
\par
Space-based coronagraphs often acquire broadband white-light polarization observations of the continuum corona (e.g. SOHO/LASCO {\citealt{LASCO1995}} and STEREO/COR {\citealt{STEREO2008}}), which can be used to infer the line of sight electron density and F corona brightness (e.g. \citealt{Woo1997}, \citealt{Habbal2001A}, \citealt{Morgan2007}), but do not provide any information about the coronal magnetic field. The planned ASPIICS coronagraph on PROBA-3 {\citep{Galano2018}} intends to observe the continuum corona between $\approx 1.1$ and 3 \Rs, while METIS/COR on Solar Orbiter {\citep{Antonucci2017}} will cover a variable distance range, where it can observe between $\approx$ 1.2 to 3.0 \Rs (1.6 to 4.1 \Rs) at minimum (maximum) orbital distance. ASPIICS and METIS should provide images of the continuum coronal emission with substantially higher resolution, lower noise and less scattered light compared to LASCO and STEREO, but will not surpass the radial extent and spatial resolution that is possible to achieve during a TSE.

\par

{Inferences of the coronal magnetic field topology are possible with} Extreme Ultra-Violet (EUV) observations {(e.g. SDO/AIA; \citealt{Lemen2012})} {as demonstrated by {\citealt{Schad2017}}.} {However, they remain} limited in spatial extent due to their collisionally excited nature (see Section 1 of \citealt{Boe2020} for an in-depth discussion). {SWAP on PROBA-2 can perform slightly better, sometimes detecting EUV emission (\ion[Fe x] 17.4 nm) as far as 2 \Rs \ {\citep{Seaton2013}}, but only in regions with a relatively high electron density ($n_e$). METIS promises to do even better, possibly reaching 3 \Rs \ or more for the \ion[He ii] 30.4 nm and \ion[H i] 121.6 nm lines {\citep{Antonucci2017}}. Even with the improvements in EUV observations, their collisionally excited nature ($\propto n_e^2$) will cause them to be strongly weighted toward seeing regions} {with higher electron density.} Observing the corona in the EUV is consequently useful for studying the magnetic loops above active regions and other structures in the low-corona ($\lesssim 1.5$ \Rs), but is not capable of {fully} revealing the magnetic field line structure for the entire corona. At present, only during a TSE is the entire corona -- and its magnetic field structure -- observable simultaneously from just above the photosphere out to at least 6 \Rs. 

\par
The first attempts to model the coronal magnetic field {were based on extrapolations of the photospheric magnetic field outward to an upper boundary, known as the `source surface', beyond which the coronal magnetic field was assumed to be radial.} Originally proposed by \cite{Altschuler1969} and \cite{Schatten1969}, these so-called Potential Field Source Surface (PFSS) models continue to evolve. {The source surface has traditionally been set at 2.5 \Rs \ from the Sun, but some more recent} models have set the source surface as low as 1.5 \Rs \ \citep{Asvestari2019} to attempt to best fit the open flux from coronal holes in the solar wind. More sophisticated models, such as Magnetohydrodynamic (MHD) models \citep{Mikic1990, Mikic1999}, have been developed in an attempt to more completely encompass the physics of the magnetized coronal plasma, {but are more computationally expensive than PFSS models}.
\par

The photospheric magnetic observations {used as the initial conditions for coronal models}, such as those by the Michelson Doppler Imager (MDI) on SOHO \citep{MDI1995} and Helioseismic and Magnetic Imager (HMI) on SDO \citep{Scherrer2012}, rely on line of sight Zeeman polarization effects. These observations are highly uncertain {for position} angles greater than $\approx 45$ to 75$^\circ$ from disk center (e.g. \citealt{Altschuler1969, Wiegelmann2017}). Synoptic maps can compensate for this angle effect with solar longitude, provided the observations from an entire solar rotation are combined. However, this process assumes that the coronal magnetic field does not change substantially over a $\approx$ 28 day rotation period. Furthermore, observations of {photospheric} magnetic fields {at the poles} have never been possible given that every solar magnetic field observatory is located near the ecliptic plane. Even Ulysses \citep{McComas1998}, which {reached a heliocentric latitude as high as $\approx 80^\circ$, did not have any remote sensing instruments} (magnetic measurements \textit{in situ} only yield the \textit{in situ} field strength and direction). {Solar Orbiter's PHI {\citep{Solanki2015}} promises to observe the polar photospheric field for the first time enabled by its inclined orbit, though it will only be able to see one pole at a time for a subset of its orbit.}

 \par
PFSS, MHD and other models are powerful ways to explore the magnetic field of the corona and the origin of the solar wind (see \citealt{Wiegelmann2017} for a coronal modeling review), but to date these approaches have been largely unconstrained due to the absence of magnetic field measurements in the corona. Even the most complex MHD models, including a combination of coronal heating processes, are not capable of perfectly recreating the shape of the solar corona when compared with TSE observations \citep{Mikic2018}. Furthermore, no model has been quantitatively compared to the eclipse observations beyond a discussion of the general location and size of streamers and coronal holes.  

\par

In this work, we quantify the topology of the coronal magnetic field using TSE observations spread over almost two complete solar activity cycles. We use an existing database of broadband white-light eclipse images (see Section \ref{Data}), and apply the Rolling Hough Transform (developed by \cite{Clark2014}, see Section \ref{RHT}) to automatically trace the coronal magnetic field line structure. A detailed accounting of the observed features of the coronal magnetic field are given in Section \ref{results}. Implications for coronal magnetic field, existing models and solar wind formation are discussed in Section \ref{disc}. Concluding remarks are given in Section \ref{conc}.
\par

\section{Data}
\label{Data}
Quantifying the structure of the magnetic field in the mid-corona (1.5 to 3 \Rs) has remained somewhat elusive despite the advancement of ground- and space-based observatories (see Section \ref{intro}), yet observing this region is crucial for understanding the origins of the solar wind. The easiest way to simultaneously observe the entire corona (from 1 to at least 6 \Rs) with high spatial resolution and strong signal continues to be during a TSE. The many years of existing TSE observations thus present a unique opportunity to quantify the magnetic field topology of the corona throughout the solar cycle, and provide improved constraints for coronal modeling. We processed archival broadband white-light total solar eclipse images taken between 2001 to 2019 to infer the magnetic field angle relative to the radial direction for all solar latitudes {(see Sections \ref{RHT} and \ref{results})}. These 14 eclipses span almost two complete solar cycles and offer a unique probe of the direction of the coronal magnetic field. 
\par
\begin{figure*}[t!]
\centering
\includegraphics[width = 4.5in]{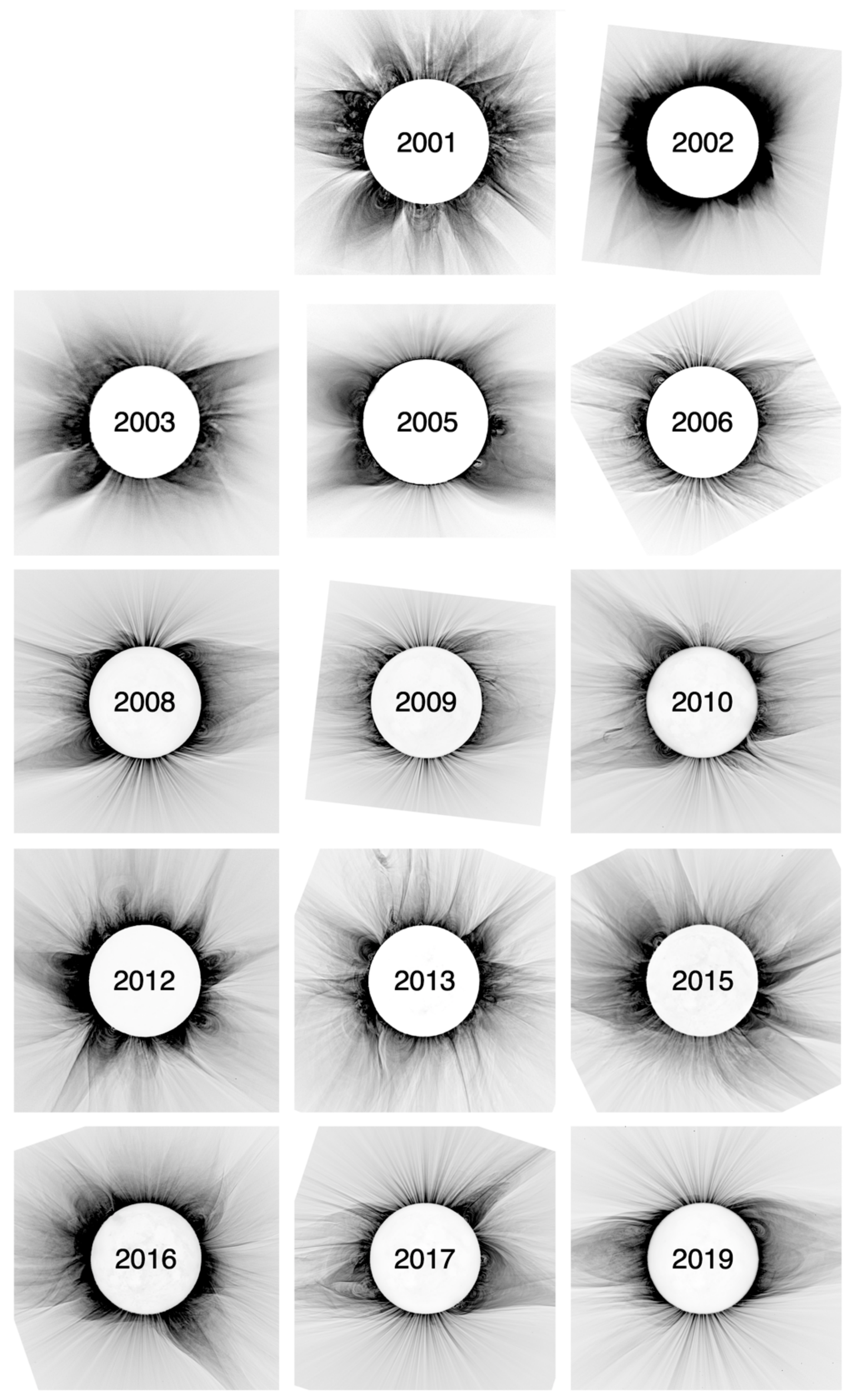}
\caption{Inverted broadband white-light images of the solar corona during total solar eclipses that were processed with the ACHF algorithm (see Section \ref{Data}). The eclipse coronae are shown in chronological order from 2001 until 2019, cropped inside a  2.5 \Rs \ square box. Solar north is oriented directly upward for all images. Note that these images are not photometrically calibrated, so the brightness of the image cannot be compared between eclipses. }
\label{fig1}
\end{figure*}
\begin{figure*}[t!]
\centering
\includegraphics[width = 6.5in]{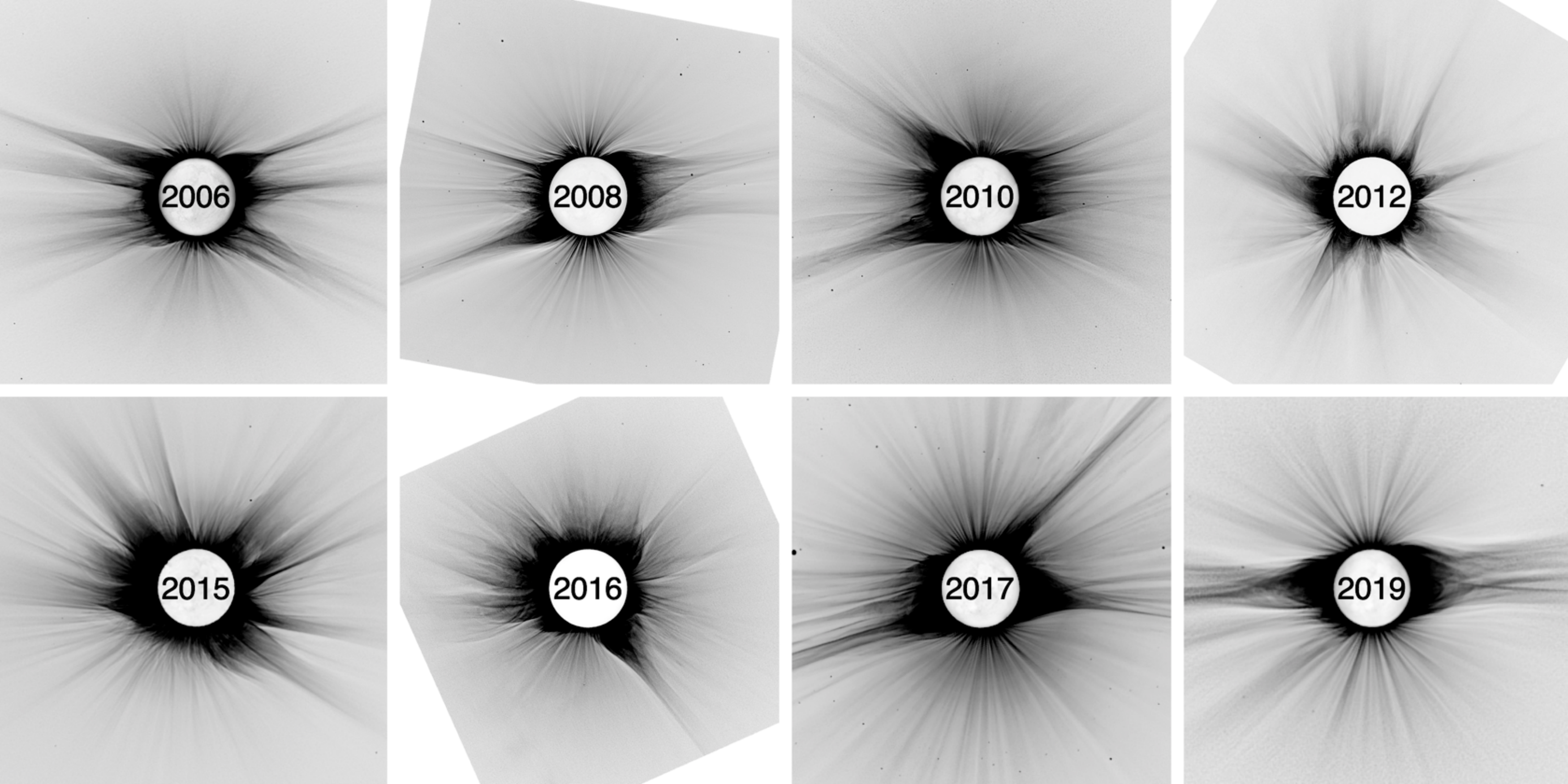}
\caption{Inverted wide field broadband white-light images of the solar corona during total solar eclipses that were processed with the ACHF algorithm (see Section \ref{Data}). The eclipse coronae are shown in chronological order from 2006 until 2019, cropped inside a 5 \Rs \ square box. Solar north is oriented directly upward for all images. Note that these images are not photometrically calibrated, so the brightness of the image cannot be compared between eclipses.}
\label{fig2}
\end{figure*}

The entirety of the data used here was taken from an open source database of processed eclipse images \footnote{a. \url{http://www.zam.fme.vutbr.cz/~druck/Eclipse/index.htm}}{, which comes from a variety of eclipse observers -- both professionals and amateurs (see Table \ref{table1})}. All eclipse data were stacked using the technique described in \cite{Druckmuller2009}, and processed with the Adaptive Circular High-pass Filter (ACHF) method \citep{Druckmuller2006} to enhance the magnetic field structure in the corona. Examples of the white light coronal data from each {of the 14 TSEs} are shown in Figure \ref{fig1}. Note that the data are not photometrically calibrated, so the intensity in the image is not representative of the relative intensity of the corona between eclipses. The processing also changes the brightness contrast between different heliocentric distances. Several of the white light images have been {previously used} to study individual eclipses (e.g. \citealt{Habbal2010a, Habbal2014, Alzate2017, Boe2018, Boe2020}). For most eclipses, multiple cameras observed the corona at different angular scales. The panels in Figure \ref{fig1} show each of the coronae {below 2.5 \Rs}. Examples from the much wider field images (inside 5 \Rs), albeit with lower resolution, are shown in Figure \ref{fig2}. 

\begin{table}[h]
\begin{center}
\begin{tabular}{|p{2.5cm}|p{4cm}|p{8cm}| }
\hline
Date & Observing Location & Observers  \\
\hline
21 Jun 2001 & Angola & \'Upice Observatory (Film camera) \\
\hline
4 Dec 2002 & South Africa & Vojtech Ru\v{s}in  (Film camera)  \\
\hline
23 Nov 2003 & Plane over Antartica & David Finlay   \\
\hline
4 Aug 2005 & Ship in Pacific Ocean & Fred Espenak \\
\hline
29 Mar 2006 & Libya & Peter Aniol, Massimiliano Lattanzi, Miloslav Druckm\"uller, Hana Druckm\"ullerov\'a \\
\hline
1 Aug 2008 & Mongolia & Peter Aniol, Miloslav Druckm\"uller, Jan Sl\'ade\v{c}ek, Vojtech Ru\v{s}in, Martin Dietzel \\
\hline
22 Jul 2009 & Marshall Islands & Peter Aniol, Miloslav Druckm\"uller, Vojtech Ru\v{s}in, L'ubom\'ir Klocok, Karel Marti\v{s}ek, Martin Dietzel \\
\hline
11 Jul 2010 & Tatakoto & Shadia Habbal, Martin Dietzel, Vojtech Ru\v{s}in \\
\hline
13 Nov 2012 & Australia & David Finlay, Constantinos Emmanouilidis, Man-To Hui, Robert Slobins \\
\hline
3 Nov 2013 & Gabon & Constantinos Emmanoulidis \\
\hline
20 Mar 2015 & Norway & Shadia Habbal, Peter Aniol, Miloslav Druckm\"uller, Pavel \v{S}tarha \\
\hline
9 Mar 2016 & Indonesia & Don Sabers, Ron Royer \\
\hline
21 Aug 2017 & USA & Shadia Habbal, Peter Aniol, Miloslav Druckm\"uller, Zuzana Druckm\"ullerov\'{a}, Jana Hoderov\v{a}, Petr \v{S}tarha \\
\hline
2 Jul 2019 & Chile & Peter Aniol, Miloslav Druckm\"uller \\
\hline
\end{tabular}
\caption{Observing dates, locations and observers for the TSE data used in this work.}
\end{center}
\label{table1}
\end{table}

\par

\par

\section{RHT Field Line Tracing Method}
\label{RHT}
Imbedded in this {TSE} dataset is a record of the coronal magnetic field {topology with high spatial resolution between 1 and over 6 \Rs, which is quantified here for the first time}. {It is important to underscore that the corona is a 3-dimensional source; all the TSE images are line-of-sight (LOS) integrations of the corona, resulting in a 2D image that has been collapsed into the plane-of-sky (POS). Since the coronal
density drops sharply with heliocentric distance, the observed intensity of the corona is substantially weighted toward the field lines that are the closest to the solar limbs in the POS (perhaps $\approx \pm 20^\circ$ of the POS). Thus, the final observables presented in this paper should be considered as a density and angle weighted average of the coronal magnetic field in the vicinity of the POS.}
\par

To quantify the coronal magnetic field structure, we apply a line tracing algorithm developed by \cite{Clark2014} called the Rolling Hough Transform (RHT). The RHT algorithm was originally used to trace the magnetic field of the Milky Way galaxy, and has been demonstrated with solar EUV data \citep{Schad2017}. Specifically, \cite{Schad2017} performed tests with the RHT algorithm indicating that it was a robust technique for detecting magnetic field line edges -- even with low signal to noise ratio data. The implementation of the RHT algorithm to the eclipse dataset is based on that described by \cite{Schad2017}.

\par
\begin{figure*}[t!]
\centering
\includegraphics[width = 7.25in]{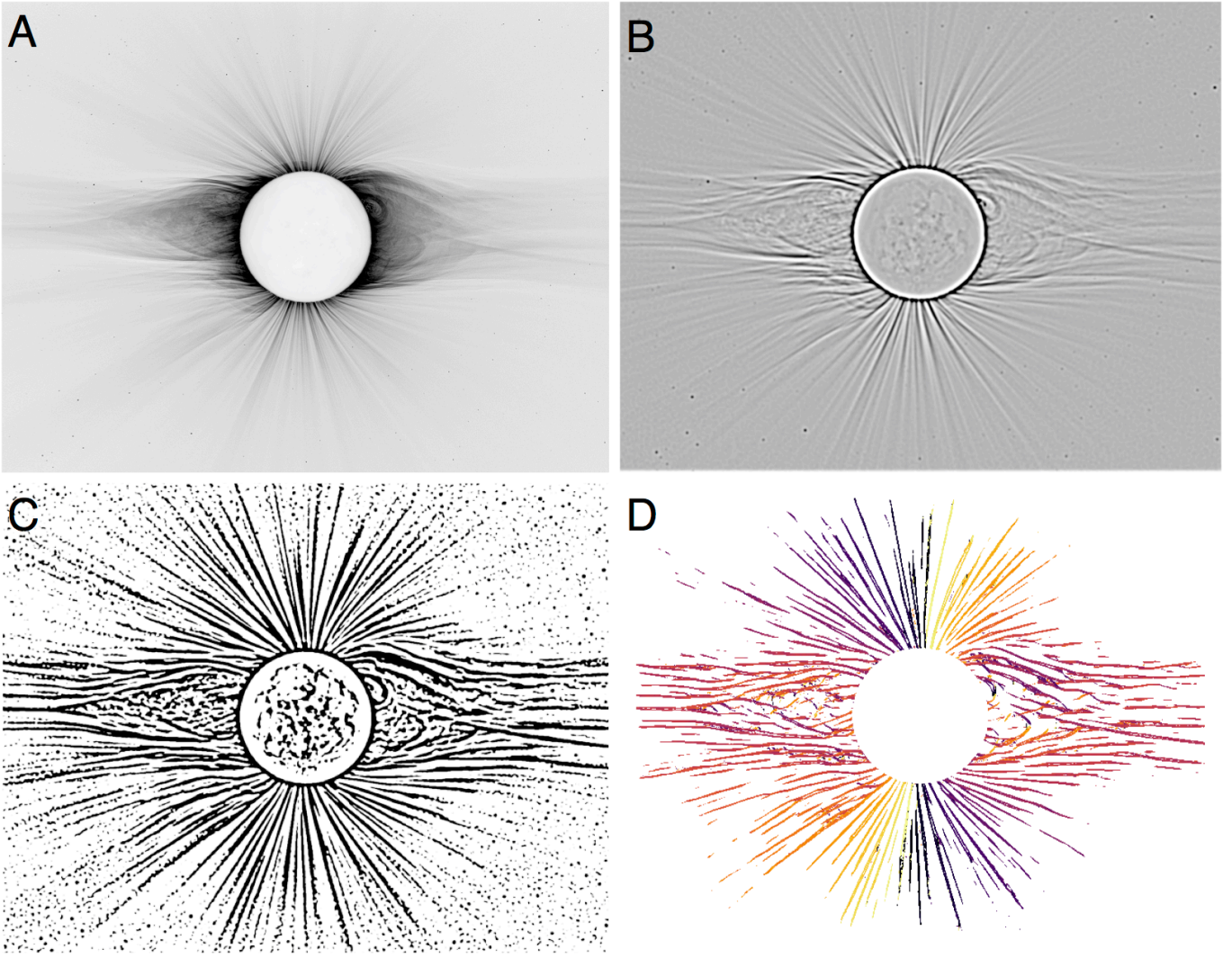}
\caption{Example of the RHT procedure using the 2019 TSE corona. The inverted ACHF processed broadband white light image (A), is filtered with a high-pass filter then blurred (B), then converted into a binary image (C). Finally the binary image is fed through the RHT algorithm to measure the angle independently for each pixel (D), {with the color scaling shown in Figure \ref{fig4}}.}
\label{fig3}
\end{figure*}

\par
\begin{figure*}[t!]
\centering
\includegraphics[width = 4.5in]{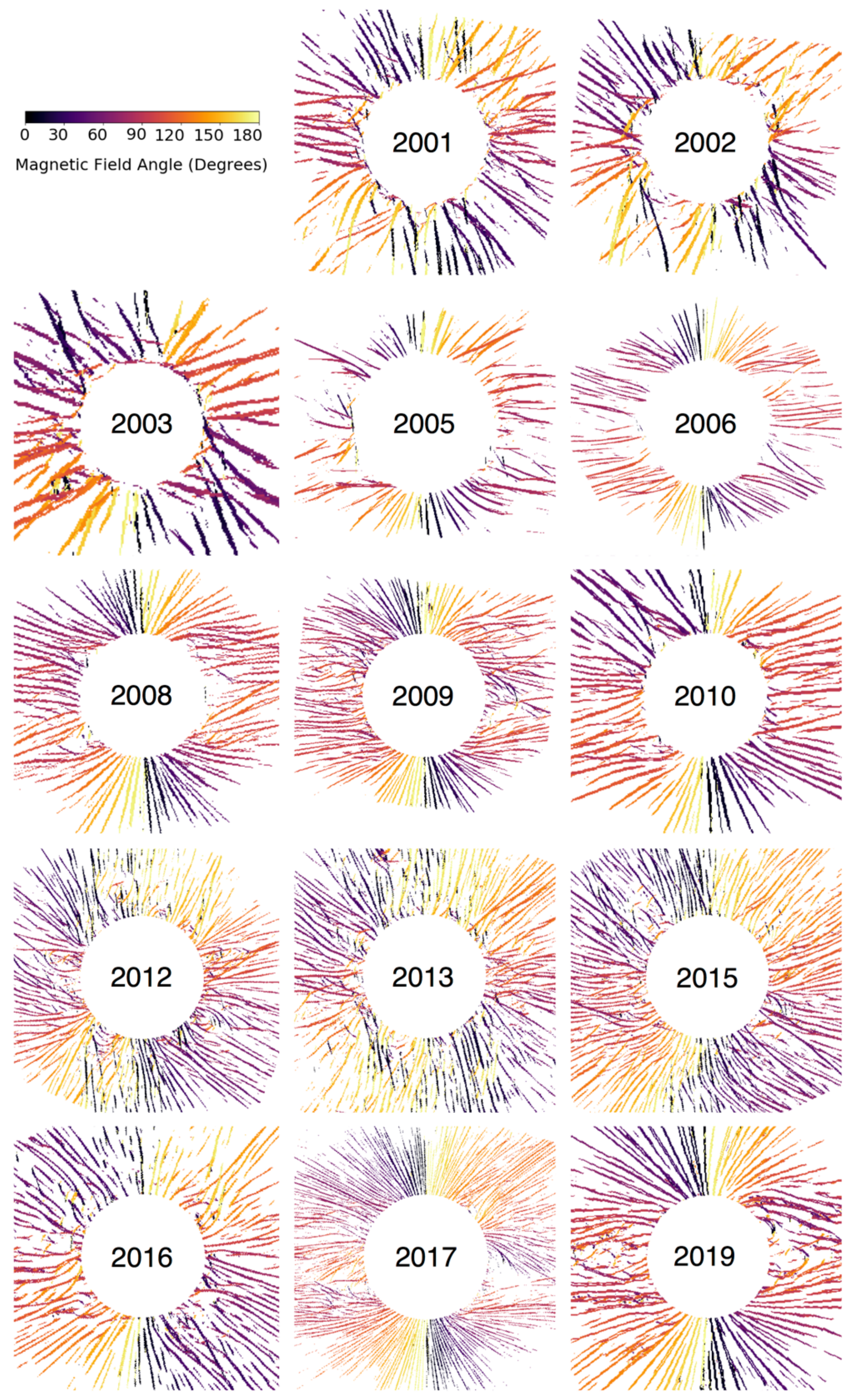}
\caption{RHT measured field angle for the same eclipse images as shown in Figure \ref{fig1}. At each pixel, the RHT algorithm reports a best fit angle with respect to the image frame. The color of the final points indicates the measured angle. The same color scaling is used for all panels, as defined in the color bar at the top left.  }
\label{fig4}
\end{figure*}
\begin{figure*}[t!]
\centering
\includegraphics[width = 6.5in]{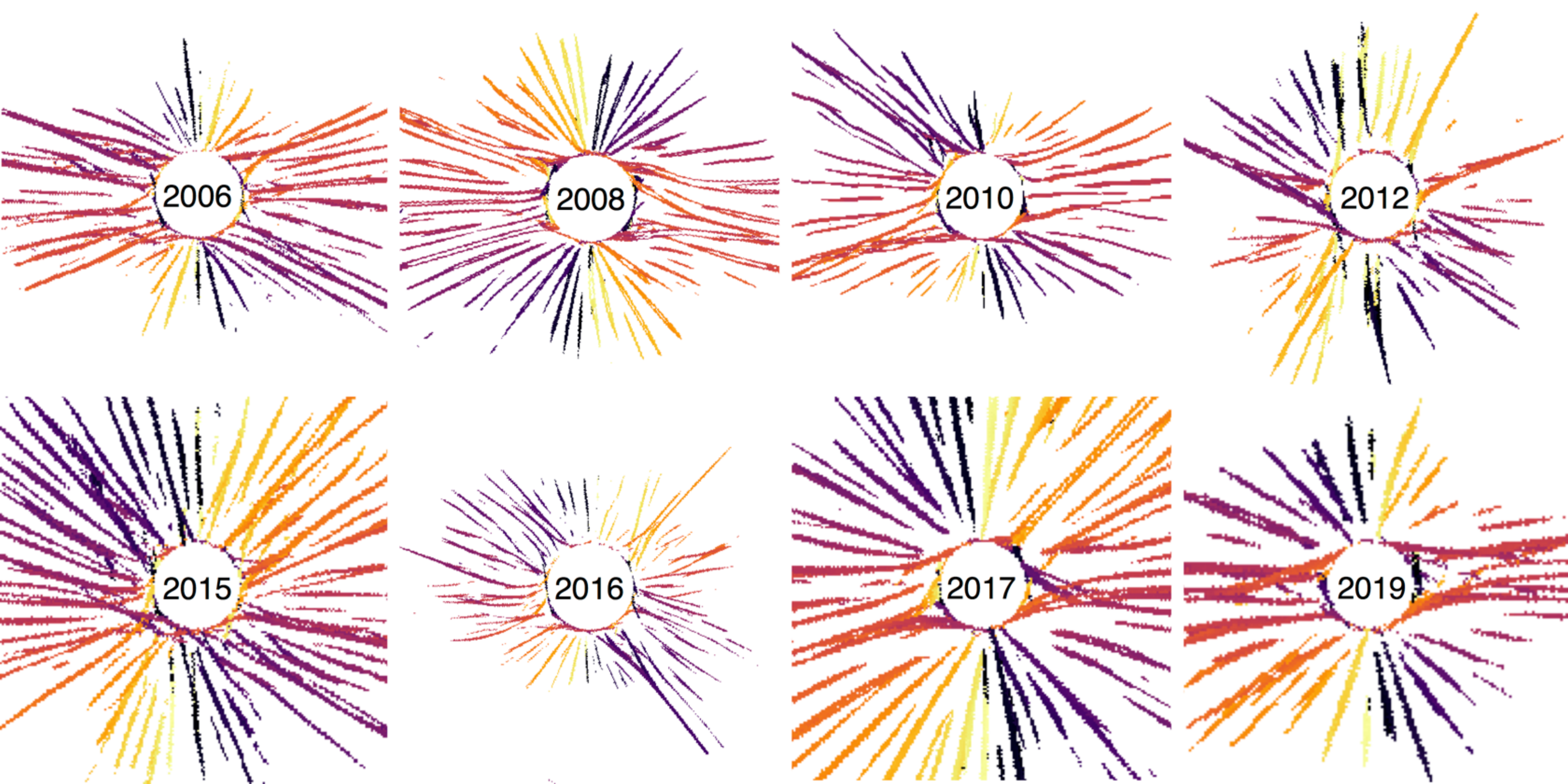}
\caption{RHT measured field angles for the wide field eclipse images shown in Figure \ref{fig2}. The exact same color scaling is used as in Figure \ref{fig4}}
\label{fig5}
\end{figure*}
\begin{figure*}[t!]
\centering
\includegraphics[width = 6.5in]{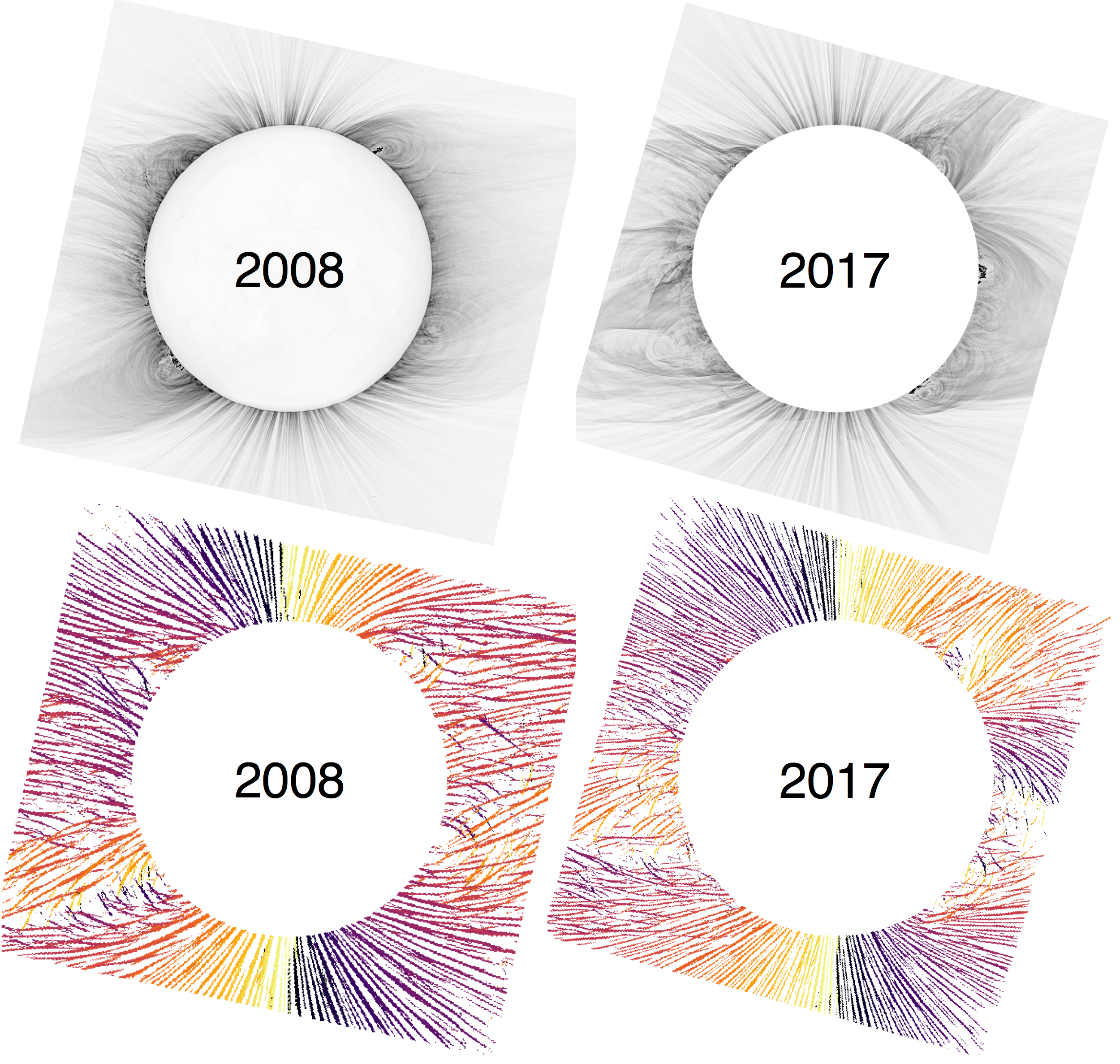}
\caption{Example of the two highest resolution coronal images that were used in this study. The top panels show the original inverted eclipse images, with the corresponding RHT measurement below. The same scaling is used on the RHT angle colors as in Figure \ref{fig4}. Fine-scale structure is visible ubiquitously throughout the corona, with {topologically `open'} field lines dominating the majority of the corona, even at 1.2 \Rs.}
\label{fig6}
\end{figure*}

First, the white light eclipse images are processed with a Gaussian high-pass filter, and then blurred at the same resolution to enhance edges in the image and eliminate any low frequency structure that is not a direct result of the magnetic field {topology}. Next the image is converted to a binary scale before finally being processed with the RHT algorithm. An example of the procedure for an image from the 2019 TSE is shown in Figure \ref{fig3}. The RHT algorithm determines the angle of the magnetic field line independently for each pixel with respect to the image coordinates {(as shown in Figs. \ref{fig3}, \ref{fig4}, \ref{fig5} and \ref{fig6}, see Fig. \ref{fig4} for the color scaling)}, by attempting to fit a line to every possible direction inside a box centered on each pixel in the binary data. It then reports the relative probability for every possible field angle direction, which is used to determine the uncertainty of each measurement, and the relative certainty between various pixels. This process is repeated independently throughout the image, so every pixel in the final image is a unique observation. The RHT algorithm does not connect features between pixels, so the lines seen in Figures \ref{fig3}, \ref{fig4}, \ref{fig5} and \ref{fig6} are not connected by the algorithm. Instead, the prevailing direction of the line is the same for the collection of pixels along the line, and thus the {continuity of the features (i.e. consistent angle/color) indicates that the measurements are robust.}

 \par

\par

\par

\par

Since the TSE data were acquired with a wide range of telescopes {and detectors} of varying quality (2001 and 2002 were done with film), each eclipse had to be manually processed to maximize the field line structure in the binary image. The Gaussian sigma of the high-pass filter ranged between $\sigma_R =$ 0.002 and 0.05 \Rs \ (in pixel units), the RHT window size ranged from $W_{RHT} = $ 0.08 to 1.0 \Rs, and the RHT smearing radius ranged between $R_{RHT} =$ 0.02 and 0.4 \Rs \ (see \cite{Clark2014} and \cite{Schad2017} for a discussion of these parameters and how they affect the final results). In general, the final measured angle by the RHT algorithm is insensitive to the exact choice of parameters (i.e. $\sigma_R$, $W_{RHT}$ and $R_{RHT}$). The main need for different parameters between eclipses is to maximize the resolution, while still capturing the {coronal field topology} to as large a distance as possible. The only problems occur when the structures are over-sampled or under-sampled. Over-sampling occurs where the size of the RHT fitting box is smaller than the width of the field line in the binary image {-- causing} the RHT algorithm to report angles perpendicular to the true field line angle. Under-sampling occurs when the RHT fitting box is too large; in that case it sees several different field lines at the same time, and thus is not capable of resolving the field line structure. These effects necessitated changing the parameters manually to maximize the potential for the RHT algorithm to measure the field line angles in the largest parameter space possible.

\par
{The initial angles returned by the RHT algorithm} are given with respect to the image frame {(as shown in Figs. \ref{fig3}, \ref{fig4}, \ref{fig5} and \ref{fig6}}). {These angles are subsequently converted to be} with respect to the radial direction. The angles relative to radial prevent any angle discontinuity effects (i.e. {0$^\circ$ vs. 180$^\circ$}) and enables a direct comparison of structures throughout the corona for a single eclipse and between eclipses. In many cases, both narrow{-field} (typically 1-3 \Rs) and wide-field (often up to {$\geq$ 6} \Rs)  coronal data are available from the same eclipse. In a few cases, there were very high resolution data of the lowest regions in the corona, which show detailed structures ubiquitously (see Figure \ref{fig6}). When multiple images are available from a single eclipse, the data are stacked to generate a continuous measurement from the lunar limb to several solar radii. All RHT data from the wide-field images below 2 \Rs \ were removed from the stacked results, due to the poor {spatial} resolution compared to the narrow-field images.

\par
\begin{figure*}[t!]
\centering
\includegraphics[width = 4.5in]{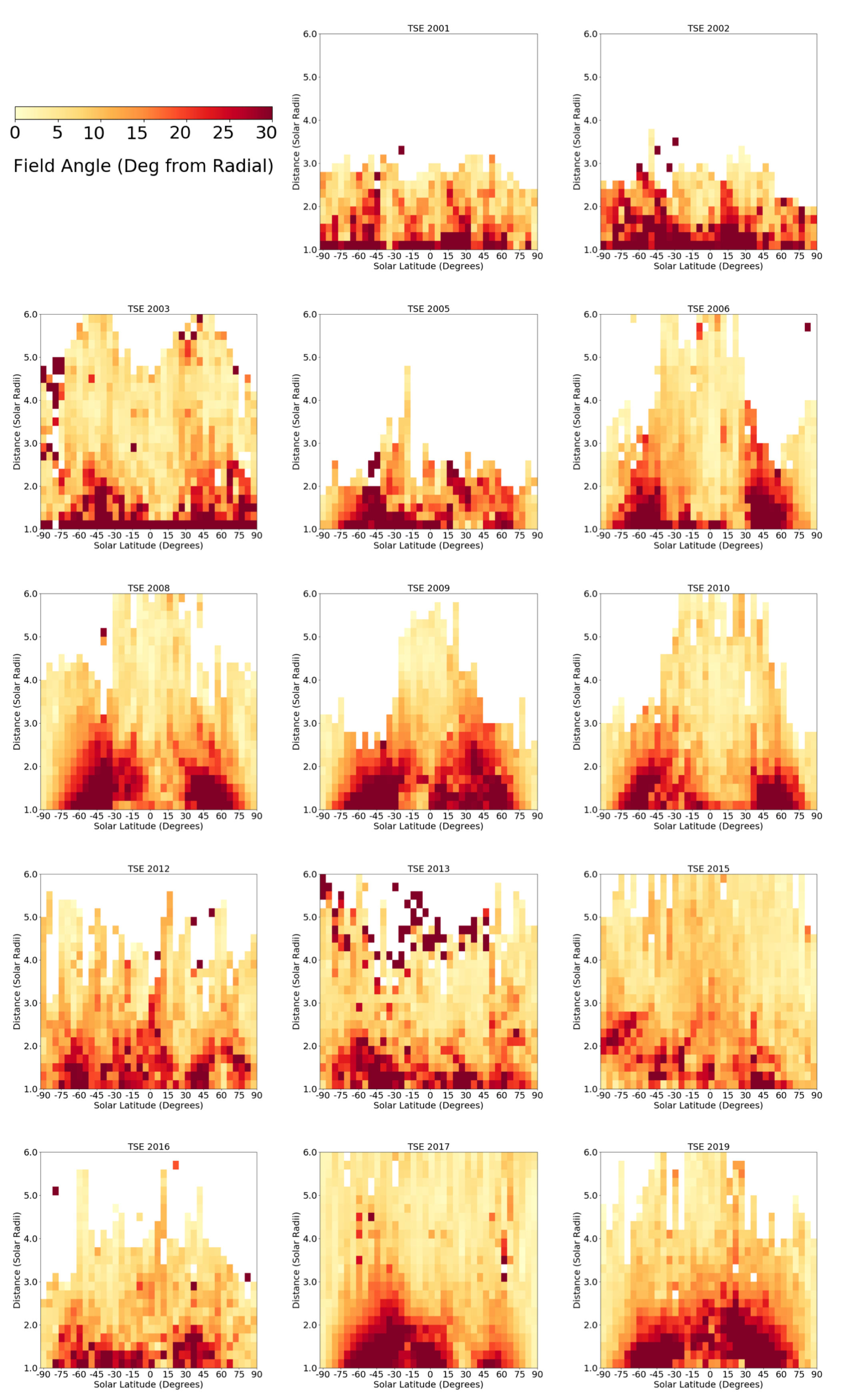}
\caption{Magnetic field angle with respect to radial for all eclipses used in this work. The RHT data is stacked {(for TSEs with multiple images available)} and binned to create the final result shown here (See Section \ref{RHT} for details).}
\label{fig7}
\end{figure*}

\par
\begin{figure*}[t!]
\centering
\includegraphics[width = 4.5in]{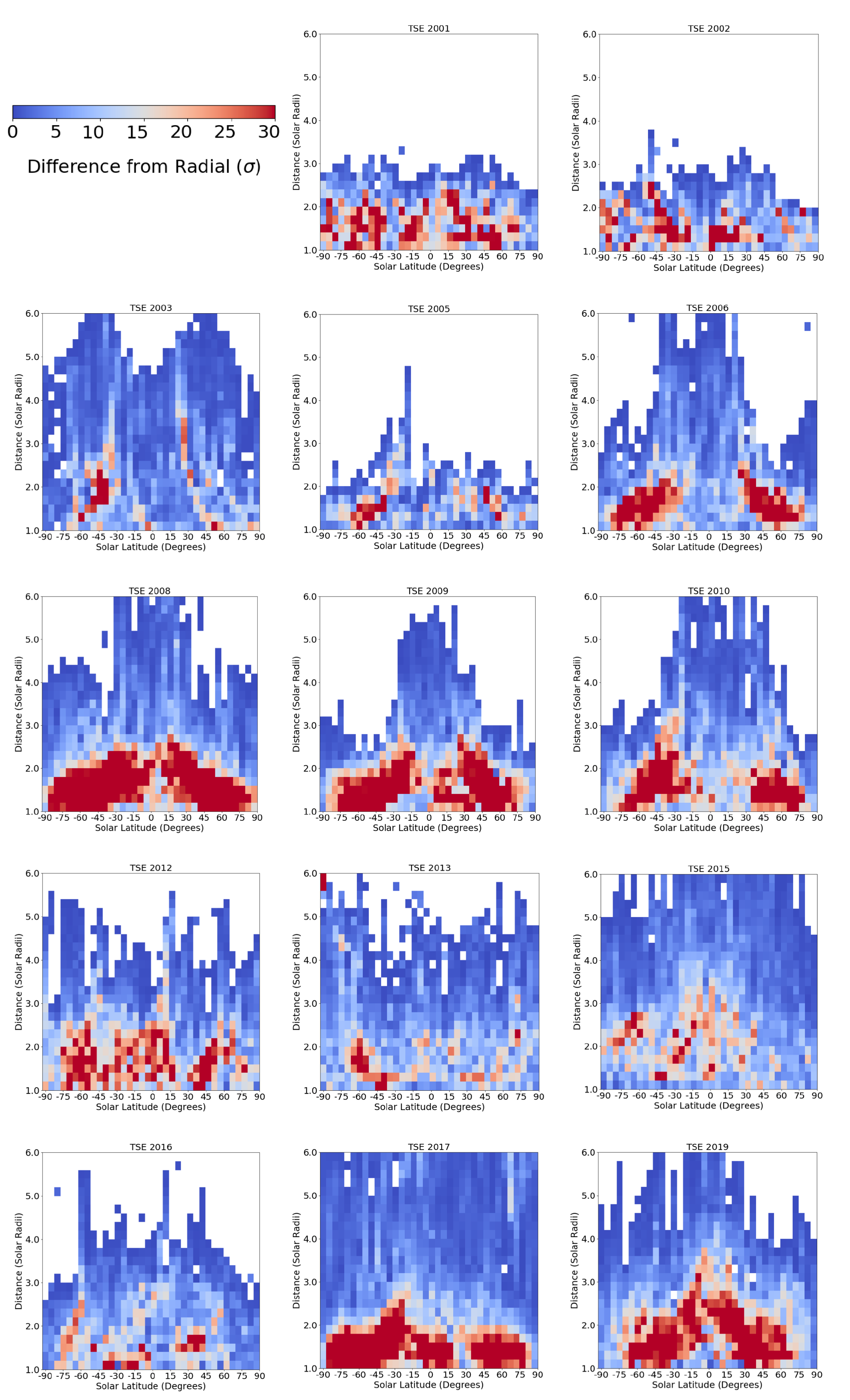}
\caption{{Statistical} significance of how far the magnetic field was from the radial direction, based on the data in Figure \ref{fig7} divided by the uncertainty of the mean field angle measurement in each bin.}
\label{fig8}
\end{figure*}

\par


\section{Results}
\label{results}

The combined data from each eclipse are shown in Figure \ref{fig7} as histograms of the field line angle with respect to the radial direction. Figure \ref{fig8} shows the same results as Figure \ref{fig7}, but in units of the relative uncertainty for the given bin (i.e. the {statistical} significance of the difference from radial). The uncertainties are propagated through the weighted mean inside each bin based on the reported RHT probabilities for each data point inside the bin. 
\par
The direction of the coronal magnetic field with respect to the radial direction, for the individual eclipses shown in Figures \ref{fig7} and \ref{fig8}, indicates a high variance in coronal structures over the solar cycle. Near solar minimum (roughly 2009 and 2019), the corona has two lobes of highly non-radial field lines between about 15 and 75$^\circ$ latitude in the North and South. The location of the non-radial lobe tends to move toward the equator with higher heliocentric distance, ending around 2.5 to 3 \Rs. Near solar maximum (roughly 2001 and 2013), small pockets of non-radial field {topology} appear throughout the corona without the large-scale lobe structure as seen at solar minimum. These observations are consistent with the concept that the coronal magnetic field is dominated by dipolar and quadrupolar magnetic moments during solar minimum, while higher order moments begin to {have a larger impact on the global coronal magnetic field} at solar maximum. {Indeed, the shape of the coronal magnetic field topology is strikingly similar for the eclipses near solar minimum, which is especially noticeable in the highest resolution TSE data (from 2008 and 2017) shown in Figure \ref{fig6}.}
\par
\begin{figure*}[t!]
\centering
\includegraphics[width = 6in]{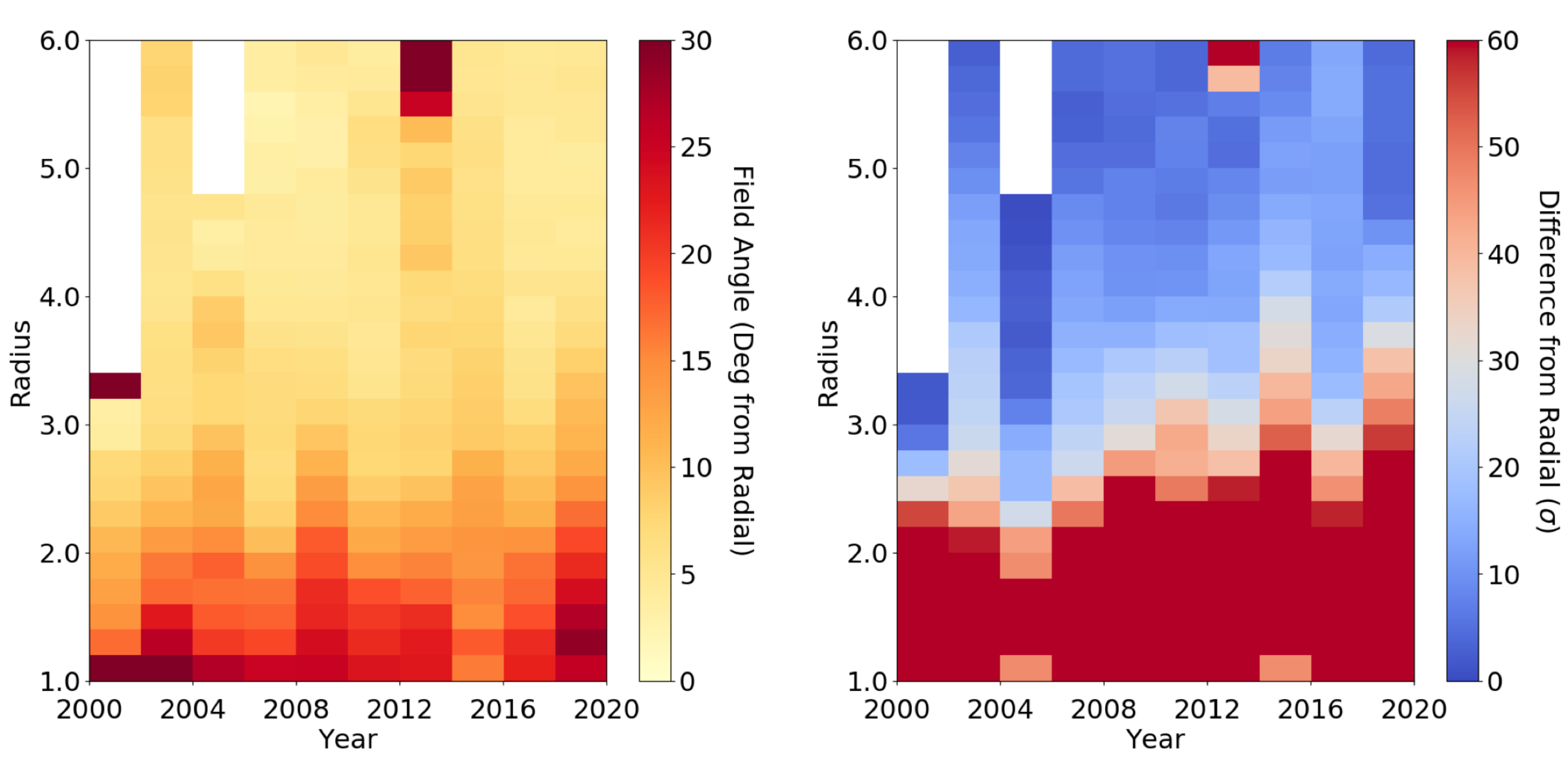}
\caption{Average magnetic field angle with respect to radial (left), and the statistical significance that the measured angle is non-radial (right), at different heliocentric distances. The TSE data is averaged over all position angles and all eclipses available for every two year period between 2000 and 2020.}
\label{fig9}
\end{figure*}

Averaging the individual eclipse data every two years across all position angles (Figure \ref{fig9}) shows the global divergence of the coronal magnetic field angle from the radial direction with the expansion of the solar wind. There does not appear to be an obvious solar cycle dependence in the average radial angle (for the entire corona), except for a slightly more non-radial angle during solar minimum in 2009 and 2019. It is clear from Figures \ref{fig7}, \ref{fig8} and \ref{fig9}, that the corona is consistently non-radial ($\geq 10^\circ$) out to 3 \Rs \ or more, to a high confidence level. The distance at which the coronal field becomes statistically consistent with radial does move slightly outward for the more recent eclipses, but this effect is due to the increased spatial resolution of the recent eclipse observations that allow for more independent measurements of the coronal magnetic field inside the same bin (thus decreasing the uncertainty). At a distance of {about} 4 \Rs \ however, the field does become almost entirely radial. One notable exception is the highly non-radial structure at 6 \Rs \ in 2013, which is due to a Coronal Mass Ejection (CME) that was previously reported by \cite{Alzate2017}.

\par
\begin{figure*}[t!]
\centering
\includegraphics[width = 6in]{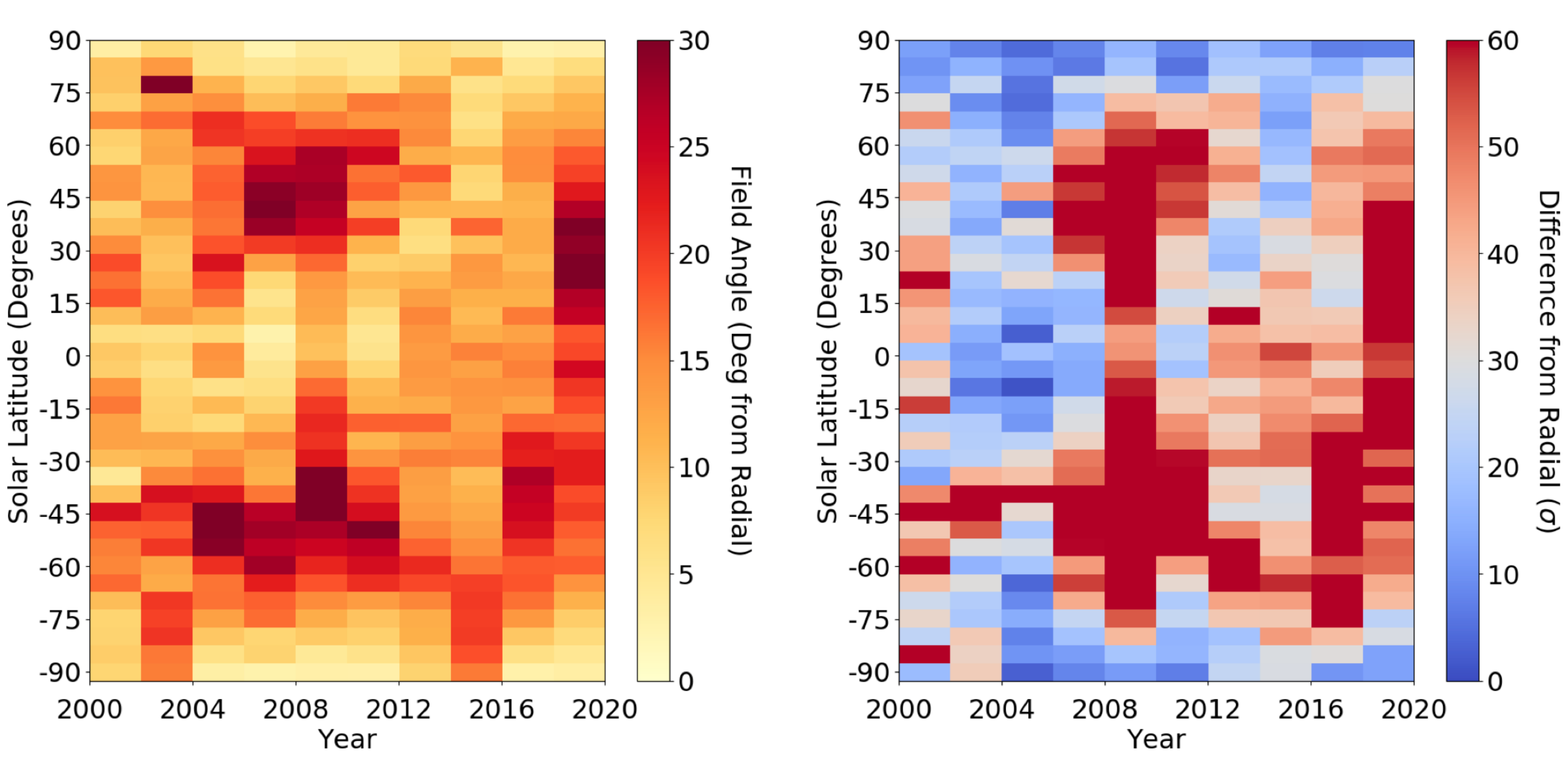}
\caption{Average magnetic field angle with respect to radial (left), and the statistical significance that the measured angle is non-radial (right) at different solar latitudes. The TSE data is averaged over a distance range of 1.5 to 3.0 \Rs \ and all eclipses available for every two year period between 2000 and 2020.}
\label{fig10}
\end{figure*}

The average deviation of the magnetic field from radial at different coronal latitudes over the {past} 20 years {of observations} is shown in Figure \ref{fig10}, where {the} field angle measurements between 1.5 and 3.0 \Rs \ are averaged for all eclipse data in two year periods. During solar minimum there are large, highly non-radial regions between 15 and 75$^\circ$ latitude, both in the north and south. While the mid-latitudes have a highly non-radial field during solar minimum, the {field lines above the poles} are almost precisely radial. During solar maximum, there are more non-radial fields throughout the entire corona, but the difference from radial is substantially smaller than in the solar minimum lobes at mid-latitudes. {Throughout the solar cycle, there exist mostly radial fields above the solar equator.}

\par

\par

\par
\begin{figure*}[t!]
\centering
\includegraphics[width = 6.75in]{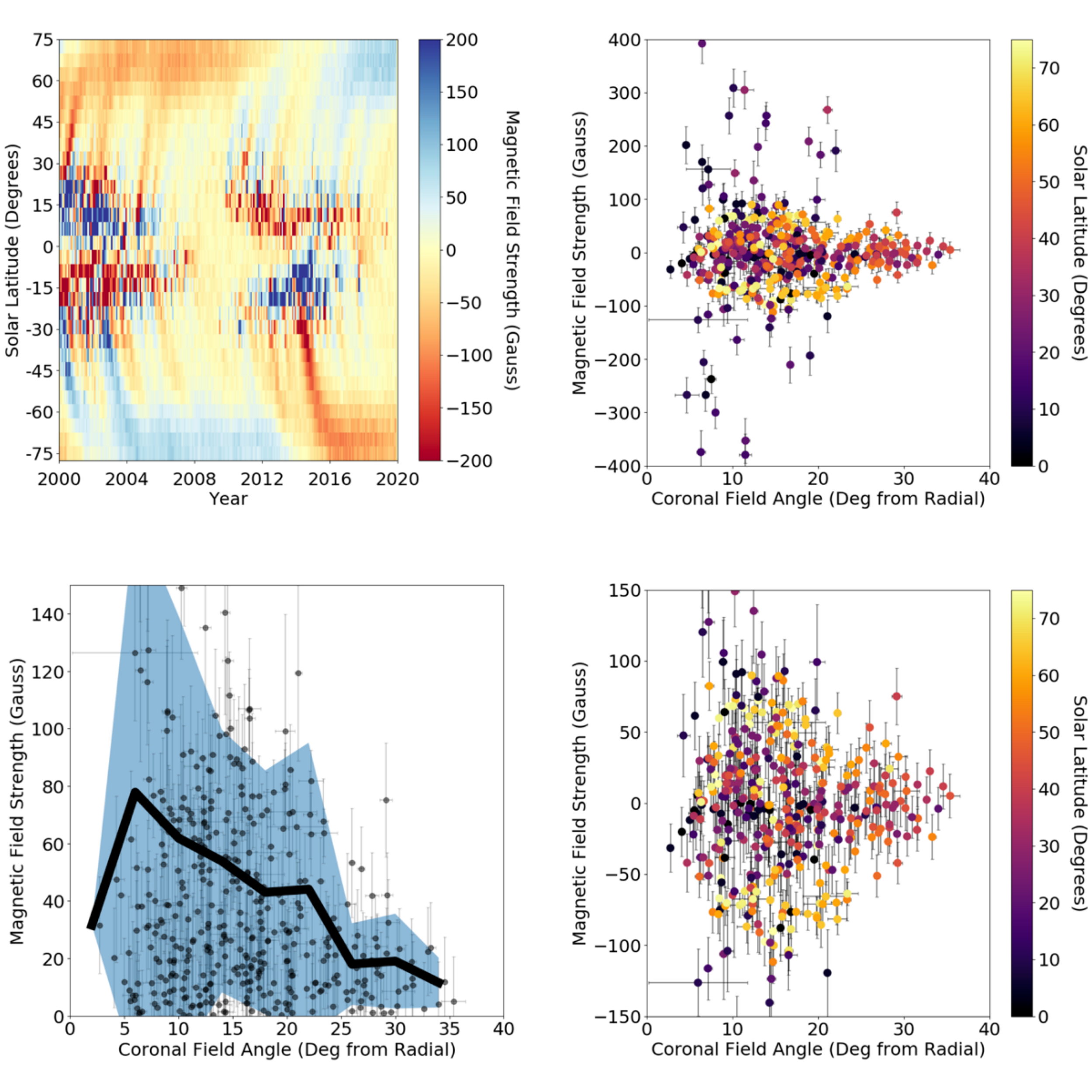}
\caption{Top left: Compilation of SOHO/MDI and SDO/HMI synoptic map magnetic field measurements averaged over all solar longitudes for each Carrington rotation, within every 5 degrees of solar latitude between -75 and 75. Top right: Average magnetic field strength in the photosphere at the same Carrington rotation at the time of each TSE, compared to the average field angle in the corona between 1.5 and 3.0 \Rs. The color of the points indicates the absolute value of the solar latitude for the data. Bottom left: Same as top right, but with the absolute value of magnetic field strength. The black line indicates the mean field strength for every 4 degrees of deviation from a radial magnetic field, and the blue region indicates the 1 $\sigma$ spread on the mean. Bottom right: Same as top right, but with decreased bounds on the magnetic field strength ($|B| < 150$ G) to show detail in the distribution.}
\label{fig11}
\end{figure*}

The coronal magnetic field {topology} is compared with the photospheric magnetic field in Figure \ref{fig11}. The photospheric magnetic field data in the top left panel of Fig \ref{fig11} are a compilation of SOHO/MDI \citep{MDI1995}  and SDO/HMI \citep{Scherrer2012} photospheric magnetic field observations. Level 2 synoptic maps were averaged in magnetic field strength at all longitudes for each Carrington rotation. For epochs where both MDI and HMI were operational, we averaged the two datasets. We then compared the photospheric magnetic field strength with the average magnetic field angle observed in the corona above, between 1.5 and 3.0 \Rs. Every latitude bin for each eclipse (as in Figs. \ref{fig7} and \ref{fig8}) is compared to the same latitude in the photospheric magnetic field for the Carrington rotation of the given eclipse. The scatter of the correlation between the photospheric field strength and the coronal field angle is shown with different scalings in the remaining panels of Figure \ref{fig11}. The top right and bottom right panels are identical except for the scaling of the magnetic field strength to show different structures in the data (see Section \ref{disc}). The bottom left panel shows the average magnetic field strength (absolute value) for different coronal field angles. The regions of the highest magnetic field curvature ($\gtrsim 20$ degrees from radial) all have a weak average magnetic field below in the photosphere, while the regions of the highest magnetic field strength in the photosphere tend to result in a much more radial coronal magnetic field.

\par

\section{Discussion}
\label{disc}
As described in Section \ref{results}, the RHT algorithm is used here to infer the coronal magnetic field topology for 14 unique eclipse coronae -- covering two solar activity cycles. These results establish that the coronal magnetic field is almost always non-radial below 4 \Rs, and that the distribution of the non-radial fields is a function of latitude and solar cycle. Next, we give an overview of the inferred topology for the coronal magnetic field and discuss the implications of these results for coronal models and solar wind formation.

\subsection{Fine-scale Structure and Coronal Radiality}
\label{radial}
\par
In all the TSE magnetic field {inferences}, very fine-scale structure is present and appears limited only by the spatial resolution of the observations. The higher resolution TSE data from the past few years show increasingly fine-scale magnetic field structure, {as illustrated by the two} cases presented in Figure \ref{fig6}. Coronagraphs have been used to previously explore the fine-scale structure of the corona, especially with regard to observing the evolution of CMEs from the corona into the heliosphere (e.g. \citealt{Gopalswamy2009}). \cite{DeForest2018} {previously} demonstrated {not only the presence of} fine-scale structures in the outer corona ($\gtrsim 3$ \Rs), {but also their dynamical evolution on short time scales. These observations are} only possible with current coronagraphs when using exceptionally long exposure {times} combined with modern image processing techniques. The eclipse data presented here indicate that the fine-scale structures are present from the outer corona region explored by \cite{DeForest2018}, through the mid-corona (1.5 to 3 \Rs) and down to the solar surface. 

\par
{Missing from current coronagraph data however, is information about the mid-corona between about 1.5 and 3 \Rs \ (see Section \ref{intro}). Insights into this data gap are provided at present by TSE observations as done here. In general, we find that the coronal magnetic field is non-radial in the low corona ($\lesssim 1.5$ \Rs) and becomes increasingly radial at higher heliocentric distances -- typically becoming entirely radial around 4 \Rs.} The coronal field is the most non-radial during solar minimum, where lobes of curved fields {($\approx 15^\circ$ up to $60^\circ$ from radial)} exist between about 15 and 75$^\circ$ latitude (see Figs. \ref{fig7}, \ref{fig8} and \ref{fig10}). The lobes are often tipped toward the equator, where the higher latitude regions become increasingly radial at lower heliocentric distances. These lobes often have a gap {of almost radial field lines} between them near the equator, which perhaps can be thought of as the signature of the beginning of the heliospheric current sheet. Likely the shape and presence of this morphology during solar minimum indicates that the corona is dominated by a dipolar and quadrupolar field component. During solar maximum, the corona is less organized; small scale non-radial structures are {ubiquitous, they} vary wildly between eclipses and do not extend to nearly as large of a heliocentric distance. 

\par
Comparing the photospheric magnetic field with the average magnetic field angle in the middle corona (see Fig. \ref{fig11}) suggests that, {on average,} strong photospheric fields  ($|B| \gtrsim 100$ G {, see Section \ref{results}}) result in more radial fields in the corona above. Clearly there are closed field lines in the {lower} corona around strong field regions with sunspots and filaments/prominences, but they {typically} do not appear to extend into the mid-corona. There can be radial fields above weak regions, but stronger photospheric fields {are correlated with} a more radial coronal field.

\subsection{Implications for PFSS and MHD Models}
\label{models}
Potential Field Source Surface (PFSS) models have long been used as a simple way to model the coronal magnetic field, using the photospheric magnetic field as a boundary condition (see Section \ref{intro}). Implicit in the model is an assumption of a source surface radius, beyond which the field is assumed to be entirely radial. The source surface is often set at 2.5 \Rs, but has been set as low as 1.5 \Rs \ (e.g. \citealt{Asvestari2019}). Our inferences of the coronal magnetic field {topology} clearly indicate that this is not a {valid} assumption for any period throughout the solar cycle. In fact, Figures \ref{fig7}, \ref{fig8}, and \ref{fig9} indicate that the coronal magnetic field is non-radial often out to 3 \Rs, with some eclipses having regions of the corona that remain non-radial to {beyond} 4 \Rs. The {topology} of the magnetic field in the middle corona (see Fig. \ref{fig10}) establishes that the deviation from the radial direction depends strongly on latitude and solar cycle. Clearly the distance at which the coronal field becomes radial is a constantly fluctuating and complex boundary between about 2 and 4 \Rs \ around the Sun. This radial distance follows a behavior similar to the shape of the freeze-in distances of Fe $10^+$ and Fe $13^+$ as observed with 2015 TSE data by \cite{Boe2018}, where the ions had lower freeze-in distances in coronal holes, compared to the streamers. This work provides the first observational evidence in support of a non-spherical source surface, as suggested by \cite{Riley2006} using MHD modeling. 
\par

{The main, and currently unavoidable, limitation of PFSS and MHD models is the necessity of photospheric magnetic field observations, which are restricted to around 75 degrees latitude of the equator -- and the need for a full rotation of data to probe all solar longitudes (this may be improved soon with Solar Orbiter, see Section \ref{intro}). Still, the most complex 3D MHD models coupled with coronal heating models (i.e. {\citealt{Mikic2018}}) are able to generate a somewhat reasonable field line structure for the middle corona, especially if the polar flux is corrected in the model to qualitatively fit a TSE image as shown by {\cite{Riley2019}}. Indeed, the choice of model parameters can strongly influence the state of the modeled plasma as well as the magnetic field topology {\citep{Lionello2009,Downs2010}}. The more traditional and lower resolution PFSS models (e.g. see {\citealt{Wiegelmann2017}}) with low source surfaces ($\leq 2.5$ \Rs) however, appear to over predict the curvature of the coronal magnetic field compared to what we have inferred in this work. It is not clear if this is an issue with insufficient model resolution, an improper selection of the source surface, or perhaps both.}
 \par
{Under-resolved structures and incorrect assumptions about the coronal magnetic field will have impacts on models of the extended corona and beyond, as PFSS and MHD models are commonly used as a tool to link solar wind observed {\textit{in situ}} back to specific sources on the Sun (e.g. {\citealt{Zhao2017}}). The magnetic field angle results presented here could be used as a constraint for future PFSS and MHD models, in combination with the photospheric field, especially to nail down the somewhat elusive high-latitude magnetic field strength (as explored by {\citealt{Riley2019}}).}

\subsection{Implications for Solar Wind Formation}
\label{wind}

{Even though the typical angle of magnetic field lines in the low corona is non-radial, many of the fine-scale structures present throughout the eclipse coronae presented in this work (see Section \ref{radial}) can be traced continuously from just above the lunar limb outward to several solar radii (see Fig. \ref{fig5}). Once a field line appears almost radial at more than $\approx 3$ or 4 \Rs, it is probably safe to assume that the field line is open and the plasma on that field line will flow the heliosphere to create solar wind. Given that the rays seen at $\approx 3-6$ \Rs \ connect to structures at almost every latitude back to the solar limb, this suggests that a large fraction of the field lines emerging from the low corona are in fact open field lines. Of course, we cannot concretely conclude that these lines are indeed open flux without a direct measurement of the magnetic field, but the inferred magnetic topology presented here seems to imply that this is the case. Such apparently `open' field lines are present seemingly regardless of the presence or absence of coronal holes on the solar disk.} Occasionally, {these `open'} field line structures are projected into the same line of sight as closed field regions, indicating that the closed field region does not dominate the entire line of sight. 
\par

{The idea that open flux may indeed be originating from what might be classically considered as a `closed' structure (e.g. an active region) has been suggested previously by {\cite{Sakao2007} and \cite{Harra2008}} using} soft X-ray {observations}. Additionally, EUV spectroscopy with EIS on Hinode {has} found Doppler-shifted outflows of 20-50 km s$^-1$ that can persist for dozens of hours above an active region \citep{Doschek2008}{. Recent work with high resolution PFSS models have shown that open flux can be rooted in and near active regions and streamers (e.g. {\citealt{Derosa2018}), and that}} the outflows at low latitude regions ($<$ 40 degrees) can explain the mass loss needed to create the slow solar wind {\citep{Brooks2015}}. {Even the fast solar wind flux cannot be fully accounted by coronal holes alone {\citep{Linker2017}}}
\par

{Recently, some studies have focused on the interaction of coronal holes adjacent to streamers and active regions (open-closed boundary) as a mechanism to explain the slow solar wind (e.g. {\citealt{Antiochos2011, Titov2011}}). These models seem to account somewhat for the complexity of the solar wind, and the presence of open flux near the solar equator, but still appear to overestimate the field line curvature (see Section \ref{models}) and underestimate the fine-scale structure seen in the TSE corona (see Section \ref{radial}) -- likely due to an insufficient model resolution. Perhaps an explicit coronal hole, or coronal hole boundary, is not required to explain the diversity of solar wind plasma characteristics. Instead, there could be regions of the quiet Sun which can in fact inject plasma into the solar wind (e.g. {\citealt{Linker2017, Riley2019}}). If that is the case, then the precise speed and composition of the solar wind {\textit{in situ}} can be explained simply by the different thermodynamical properties of different regions in the low corona {\citep{Habbal2001A, Xu2015}}.}

\section{Conclusions}
\label{conc}
\par

{Application of the Rolling Hough Transform (RHT) to white-light TSE images, as done here for 14 eclipses between 2001 and 2019, has demonstrated that TSE observations continue to yield unique opportunities to quantify the topology of the coronal magnetic field throughout the solar cycle -- thus providing valuable constraints for coronal modeling and solar wind formation. In particular, this work showed that:

\begin{enumerate}
\item The coronal magnetic field is highly structured {down to the spatial resolution of the images, with topologically `open' field line structures visible at most latitudes throughout the corona for all solar eclipse images.}
\item The coronal field can evolve significantly in the mid-corona (1.5 to 3 \Rs) between different latitude regions over the solar cycle, often remaining non-radial out to about 4 \Rs.  
\item The most non-radial regions {in the mid-corona occur above the weakest photospheric magnetic fields, whereas regions with the strongest fields are correlated with} nearly radial {field lines.}
\item These TSE observations yield essential constraints for coronal models. We find that the canonical source surface of 2.5 \Rs \ in PFSS models is not far enough from the Sun, and that the source surface is non-spherical. 
\end{enumerate}

\par
 The RHT algorithm and procedure presented here could be used in the future with coronagraphic observations, especially with the upcoming METIS and ASPIICS coronagraphs. Inferring the coronal magnetic field topology may consequently be possible for a much larger and almost continuous dataset. The RHT tracing method could be useful for mapping the structures of CMEs in the corona and as a tool to empirically link {\textit{in situ}} solar wind observations to sources on the Sun, without any assumptions whatsoever on the origin of the structure from just above the photosphere to at least the source surface region.}

\subsection*{Acknowledgments}
This work would not have been possible without the many years of hard work and dedication of all the eclipse observers, given in Table \ref{table1}. Special thanks to Xudong Sun for suggesting the RHT algorithm for use with eclipse data. {We thank the reviewer for their exceptionally fair and insightful comments.} Financial support was provided to B. Boe by AURA/NSO with grant N97991C to the Institute for Astronomy at the University of Hawaii. NASA grant NNX17AH69G and NSF grants AGS-1358239, AGS-1255894, and AST-1733542 to the Institute for Astronomy of the University of Hawaii. M. Druckm\"uller was supported by the Grant Agency of Brno University of Technology, project No. FSI-S-20-6187.

\bibliographystyle{apj}
\bibliography{Boe2020b_arxiv}
\end{document}